\documentclass[12pt]{article}
\usepackage{amsmath,enumerate,amsfonts,color}

\usepackage{graphicx}
\usepackage{amssymb}
\usepackage{amsmath}
\usepackage{srcltx}
\usepackage{color}
\usepackage{cite}
\usepackage{url}
\usepackage{ntheorem}
\usepackage{psfrag}

\newcommand{\beadl}[1]{\begin{deqarr}\label{#1}}
\newcommand{\eeadl}[1]{\arrlabel{#1}\end{deqarr}}
\newcommand{\eead}[1]{\end{deqarr}}

\newcommand{\Yfield}{\eta}
\newcommand{\kfield}{\xi}

\newcommand{\tchi}{{\tilde \chi}}

\newtheorem{Theorem} {\sc  Theorem\rm} [section]

\newtheorem{Proposition} [Theorem] {\sc  Proposition\rm}

\theorembodyfont{\upshape}
\newtheorem{Remark}[Theorem]{\sc  Remark\rm}

\theoremsymbol{\ensuremath{\diamondsuit}}
\newcommand{\fcoco}{\small}
\theorembodyfont{\fcoco}\theoremseparator{}\theoremindent0.5cm

\theoremstyle{nonumberplain}\theorembodyfont{\fcoco}
\theoremseparator{}\theoremindent0.5cm

\DeclareFontFamily{OT1}{rsfs}{}
\DeclareFontShape{OT1}{rsfs}{m}{n}{ <-7> rsfs5 <7-10> rsfs7 <10-> rsfs10}{}
\DeclareMathAlphabet{\mycal}{OT1}{rsfs}{m}{n}

{\catcode `\@=11 \global\let\AddToReset=\@addtoreset}
\AddToReset{equation}{section}

\newcounter{mnotecount}[section]

\AddToReset{figure}{section}
\renewcommand{\themnotecount}{\thesection.\arabic{mnotecount}}

\newcommand{\mnotex}[1]
{\protect{\stepcounter{mnotecount}}$^{\mbox{\footnotesize
$
\bullet$\themnotecount}}$ \marginpar{
\ragged\hr ight\tiny\em
$\!\!\!\!\!\!\,\bullet$\themnotecount: #1} }

\newcommand{\ptccheck}[1]{\mnote{ptcchecked:#1}}
\renewcommand{\ptccheck}[1]{\mnote{\checkmark (ptc #1)}}

\newcommand{\tilr}{\tilde r}%

\newcommand{\jlcax}[1]{}
\newcommand{\eean}{\nonumber\end{eqnarray}}

\newcommand{\Uone}{\mathrm{U(1) }}%

\newcommand{\ttheta}{\tilde\theta}

\newcommand{\kk}[1]{}

\newcommand{\Span}{\mathrm{Span}}

\newcommand{\beq}{\begin{equation}}

\newcommand{\FS}       
                  {F}

\newcommand{\HS} 
       {H_{\mbox{\scriptsize volume}}}

{\ptc{this should be removed in the oberwolfach version}}%

\newcommand{\mcA}{{\mycal A}}%
\newcommand{\mcE}{{\mycal E}}%

\newcommand{\eeal}[1]{\label{#1}\end{eqnarray}}
\newcommand{\bed}{\begin{deqarr}}
\newcommand{\eed}{\end{deqarr}}
\newcommand{\bedl}[1]{\begin{deqarr}\label{#1}}
\newcommand{\eedl}[2]{\arrlabel{#1}\label{#2}\end{deqarr}}

\newcommand{\mcO}{{\mycal O}}

\newcommand{\bel}[1]{\begin{equation}\label{#1}}
\newcommand{\bea}{\begin{eqnarray}}
\newcommand{\bean}{\begin{eqnarray}\nonumber}
\newcommand{\beal}[1]{\begin{eqnarray}\label{#1}}
\newcommand{\eea}{\end{eqnarray}}

\newcommand{\be}{\begin{equation}}
\newcommand{\eeq}{\end{equation}}
\newcommand{\ee}{\end{equation}}
\newcommand{\beqa}{\begin{eqnarray}}
\newcommand{\eeqa}{\end{eqnarray}}
\newcommand{\beqan}{\begin{eqnarray*}}
\newcommand{\eeqan}{\end{eqnarray*}}
\newcommand{\ba}{\begin{array}}
\newcommand{\ea}{\end{array}}

\newcommand{\mcM}{{\mycal M}}

\newcommand{\mnote}[1]
{\protect{\stepcounter{mnotecount}}$^{\mbox{\footnotesize
$
\bullet$\themnotecount}}$ \marginpar{
\raggedright\tiny\em
$\!\!\!\!\!\!\,\bullet$\themnotecount: #1} }

\newcommand{\warn}[1]
{\protect{\stepcounter{mnotecount}}$^{\mbox{\footnotesize
$
\bullet$\themnotecount}}$ \marginpar{
\raggedright\tiny\em
$\!\!\!\!\!\!\,\bullet$\themnotecount: {\bf Warning:} #1} }

\newcommand{\R}{\mathbb R}

\newcommand{\eq}[1]{(\ref{#1})}

\newcommand{\Mext}{M_\ext}
\newcommand{\ext}{\mathrm{ext}}

\newcommand{\ptc}[1]{\mnote{{\bf ptc:}#1}}

\newcommand{\beqar}{\begin{deqarr}}
\newcommand{\eeqar}{\end{deqarr}}

\newcommand{\beaa}{\begin{eqnarray*}}
\newcommand{\eeaa}{\end{eqnarray*}}

\newcommand{\mcB}{{\mycal B}}

\newcommand{\doc}{\langle\langle \mcM\rangle\rangle}
\newcommand{\pdoc}{\partial\doc}

\newcommand{\trho}{{\tilde \rho}}

\newcommand{\hr}{{\hat r}}
\newcommand{\hx}{{\hat x}}
\newcommand{\hy}{{\hat y}}

\newcommand{\tvarphi}{{\tilde\varphi}}

\renewcommand{\ptccheck}[1]{}

\renewcommand{\Mext}{\mycal{M}_{\ext}}
\renewcommand{\doc}{\langle\langle \Mext\rangle\rangle}

\def\RR{{\mathbb R}}

\def\FS{{\mathfrak S}}
\def\CA{{\mathcal A}}
\def\CF{{\mathcal F}}

\newcommand\xke{X_{\mbox{\scriptsize Kerr}}}
\newcommand\yke{Y_{\mbox{\scriptsize Kerr}}}
\newcommand\uke{U_{\mbox{\scriptsize KN}}}
\newcommand\vke{V_{\mbox{\scriptsize KN}}}

\newcommand\chie{\chi^{\mbox{\scriptsize e}}}

\newcommand\chim{\chi^{\mbox{\scriptsize m}}}

\newcommand\chieke{\chi^{\mbox{\scriptsize e}}_{\mbox{\scriptsize KN}}}
\newcommand\chimke{\chi^{\mbox{\scriptsize m}}_{\mbox{\scriptsize KN}}}

\newcommand\mzero{m_0}
\newcommand\azero{a_0}

\def\eproof{\hfill$\square$\medskip}

\newcounter{marnote}

\newcommand{\matrixform}[2]{\left[\begin{array}{#1}#2\end{array}\right]}

\newcommand{\tilx}{{\tilde x}}

\renewcommand{\tvarphi}{\varphi}

\begin{document}

\title{A uniqueness theorem for degenerate Kerr-Newman black holes}
\author{Piotr T. Chru\'{s}ciel\thanks{LMPT, F\'ed\'eration Denis Poisson, Tours; Mathematical
Institute and Hertford College, Oxford}$\ $ and Luc
Nguyen\thanks{OxPDE, Mathematical Institute, Oxford}}
\maketitle

\begin{abstract}
We show that the domains of dependence of stationary,
$I^+$-regular, analytic, electrovacuum space-times with a
connected, non-empty, rotating, degenerate event horizon arise
from  Kerr-Newman space-times.
\end{abstract}

\section{Introduction}

A classical problem in general relativity is that of
classification of domains of outer communication of suitably
regular black hole space-times. A complete solution for
stationary, $I^+$-regular, analytic, vacuum, connected
\emph{non-degenerate} black holes has been given
in~\cite{ChCo}, building on the fundamental work
in~\cite{RobinsonKerr,Sudarsky:wald,CarterlesHouches,Weinstein1,Ha1}
and others; see~\cite{AIK,AIK2} for some progress towards
removing the hypothesis of analyticity. The analysis
in~\cite{ChCo} has been extended to the electrovacuum case
in~\cite{Costaelvac,CostaPhD} (see~\cite{Mazur,Bu,CarterCMP}
for previous results). The aim of this work is to remove the
condition of non-degeneracy in the \emph{rotating} case  (here
$\doc$ denotes the domain of outer communications; the reader
is referred to~\cite{ChCo} for terminology and further
notation):

\begin{Theorem}
 \label{Tmain30I.1}
Let $(\mcM,g)$ be a stationary, $I^+$-regular, analytic,
electrovacuum space-time with connected, non-empty,  rotating,
degenerate future event horizon $I^+(\Mext)\cap \partial \doc$.
Then $(\doc,g)$ is isometrically diffeomorphic to the domain of
outer communications of a Kerr-Newman space-time.
\end{Theorem}

\emph{Non}-rotating, degenerate, \emph{vacuum} and suitably
well-behaved solutions are expected \emph{not} to exist; here
one should keep in mind that while the usual staticity argument
for non-rotating configurations applies both for
non-degenerate~\cite{Sudarsky:wald} (compare~\cite[end of
Section~7]{ChCo}) and degenerate~\cite[Section~5]{CRTIWP}
configurations, it requires existence of a maximal surface,
which has only been proved in the non-degenerate case so
far~\cite{ChWald}. Static electrovacuum solutions with
degenerate components have been classified in~\cite{CT}, see
also~\cite{CRTIWP}.

The first element needed to prove Theorem~\ref{Tmain30I.1}, and
missing in the arguments given in~\cite{ChCo,Costaelvac} under
the current assumptions, is the \emph{global} reduction to a
harmonic map problem; equivalently, one needs to prove that the
area density of the orbits of the isometry group can be used as
one of global coordinates on the domain of outer
communications; this is established below in
Theorem~\ref{T30I.1}. The other missing element is the proof
that the harmonic map associated to $(\mcM,g)$ lies a finite
distance to a Kerr-Newman one; we do this below in
Theorem~\ref{Tfd} in vacuum and Theorem~\ref{Tfd-ev} in
electrovacuum. The remaining arguments of the proof of
Theorem~\ref{Tmain30I.1} are as in~\cite{ChCo,Costaelvac}; for
the convenience of the reader we present a few more essential
steps in Section~\ref{Sproof}.

Our analysis below can be used to provide a uniqueness theorem
for stationary \emph{and} axisymmetric space-times with several
black hole components, along the lines of Corollary~6.3
of~\cite{ChCo}; note that many such \emph{vacuum}
configurations are excluded by the analysis in
\cite{HennigNeugebauer}.

\section{Adapted coordinates}
 \label{SnhsXII.1}

Assuming $I^+$-regularity and analyticity, it follows from the
Structure Theorem~4.5 in~\cite{ChCo} that Hawking's rigidity
theorem~\cite[Theorem~4.13]{ChCo} applies, and so for each
rotating connected component of  the future event horizon
$I^+(\Mext)\cap\pdoc$ there exists on $\overline{\Mext}$ a
Killing vector field $\xi$ tangent to the generators, without
zeros on $I^+(\Mext)\cap \pdoc$, as well as a second Killing
vector field $\Yfield$, commuting with $\xi$, and generating a
$\Uone$ action on $\mcM$. Introducing null Gaussian
coordinates~\cite{VinceJimcompactCauchyCMP} near a connected
degenerate component of $I^+(\Mext)\cap\pdoc$,   the metric
there takes the form
\bel{10XII.1}
 g = -\tilr ^2\,\tilde F(\tilr ,\tilx)\,dv^2 -2dv\,d\tilr  +
 2\tilr\,h_a(\tilr ,\tilx)\,dv\,d\tilx^a + h_{ab}(\tilr ,\tilx)\,d\tilx^a\,d\tilx^b
 \;,
\ee
where $\xi=\partial_v$, the horizon is at $\tilr =0$, we write
$\tilx=(\tilx^a)$, and the $\tilx^a$'s are coordinates on a two
dimensional cross section of the horizon, which is spherical by
the topology theorem~\cite{ChWald1}. All functions are smooth
functions of their arguments near $\tilr=0$.

It has been shown in~\cite{Hajicek3Remarks}, and rediscovered
in~\cite{LP1} (see also~\cite{FiguerasLucietti}), that, for
axisymmetric stationary vacuum metrics, the leading order
behaviour of the functions above coincides with that of the
extreme Kerr metric. We choose the coordinates $\tilx^a$ at
$\tilr =0$ to coincide with the spherical Boyer-Lindquist
coordinates $(\ttheta,\tvarphi)$ of the Kerr metric
(compare~\eq{r=0-A}-\eq{r=0-alpha} below). A similar procedure
applies to the electrovacuum situation, using~\cite{LP1}. We
will return to the details of those constructions in
Sections~\ref{ssKerr} and \ref{ssKerrNewmanme}. The coordinates on the
horizon are then propagated away from the horizon so as to
obtain the form \eq{10XII.1} of the metric. Then
$\eta=\partial_\tvarphi$, and since  the commutator $[\xi,
\eta]$ vanishes, the construction leading to \eq{10XII.1} can
be carried-out so that all metric functions are independent of
both $v$ and $\tvarphi$.

It turns out to be convenient to rewrite \eq{10XII.1} as
\bean
 g &=& -\tilr^2\,F(\tilr,\ttheta)\,dv^2 -2\,dv\,\big(d\tilr
+ \tilr \lambda( \tilr,\ttheta)  \,d\ttheta\big)
+h_{{\tvarphi}{\tvarphi}}(d{\tvarphi} + \tilr
\,\alpha(\tilr,\ttheta)\,dv)^2
 \nonumber
\\
&&	+ 2h_{{\tvarphi}\ttheta}\,(d{\tvarphi}
+\tilr\,\alpha(\tilr,\ttheta)\,dv)\,d\ttheta +
h_{\ttheta\ttheta}\,d\ttheta^2\;.
 \label{10XII.5}
\eea
To obtain this form of the metric one defines $\alpha$ as
\bel{19XII.1}
 \alpha := \frac {g_{\tvarphi v }}{\tilr g_{\tvarphi \tvarphi}}
  \equiv \frac {g({\partial_\tvarphi, \partial_ v })}{\tilr g({\partial_\tvarphi,\partial_ \tvarphi})}
 \;,
\ee
and the other functions in \eq{10XII.5} are then obtained by
redefinitions:
\bel{25I.1}
 F= \tilde F - g_{\tvarphi \tvarphi} \alpha^2\;, \qquad
 \lambda = -h_{\ttheta} + g_{\tvarphi \ttheta} \alpha
 \;,
\ee
with $h_{\tvarphi\tvarphi}=g_{\tvarphi\tvarphi}$, etc. Since
$g_{\tvarphi\tvarphi}$ vanishes at zeros of $\tvarphi$,
smoothness of $\alpha$ at the zero-set of $\partial_{\tvarphi}$
requires justification; this proceeds as follows:

Since $\partial_{\tvarphi}$ and $\partial_v$ are Killing vector
fields, both $ g(\partial_\tvarphi,\partial_v)$ and $
g(\partial_\tvarphi,\partial_\tvarphi)$, and hence their ratio,
are scalar functions on space-time. So smoothness of the ratio
is obvious away from zeros of
$g(\partial_\tvarphi,\partial_\tvarphi)$. To proceed further,
we need to understand the nature of the zero-set of
$\partial_\tvarphi$.

In the current coordinate system the Killing vector field
$\Yfield$ coincides with $
\partial_{\tvarphi}$.
It is well known that a periodic Killing vector field cannot be
null on a causal domain of outer communications $\doc$. It
further follows from~\cite[Theorem~4.5]{ChCo} that in
$I^+$-regular space-times the Killing vector field $\Yfield $
cannot be null on $I^+(\doc)\cap
\partial\doc$. So, under the hypothesis of $I^+$-regularity, on
this last region the function
$g(\partial_\tvarphi,\partial_\tvarphi)$ vanishes only at zeros
of $\Yfield $. Consider then a point $p$ at which a Killing
vector field $\Yfield $ vanishes. It is also well known (see,
e.g., \cite[Proposition~7.1]{Chhighdim}) that, in four
dimensional space-times, there exists a normal coordinate
system $\{x^\mu\}$ centred at $p$ such that:
 \begin{enumerate}
\item either there exist constants $\beta_\mu\in \R$,
    $\mu=0,1$, not both zero, such that
\bel{Xfcase} \Yfield  = \beta_0
(x^0\partial_1+x^1\partial_0) + \beta_1
(x^{3}\partial_{2}-x^{2}\partial_{3})    \;;
\ee
\item or there exists a constant  $a\in \R^*$  such that
\bel{Xfcase2} \Yfield  = a \left(
(x^0-x^2)\partial_1+x^1(\partial_0+\partial_2) \right)  \;.
\ee
\end{enumerate}

Exponentiating, in normal coordinates near $p$ the action of
the isometry group $\phi_t$ generated by $\Yfield $ is linear
and takes the form
\bel{29I.1}
 \left( \begin{array}{cccc}
                  \cosh(\beta_0t) & \sinh(\beta_0 t) & 0 & 0 \\
                  \sinh(\beta_0 t) & \cosh(\beta_0 t) & 0 & 0 \\
                  0 & 0 & \cos(\beta_1 t) & -\sin(\beta_1 t) \\
                  0 & 0 & \sin(\beta_1 t) & \cos(\beta_1 t)
                \end{array}
                \right)
                \left( \begin{array}{c}
                         x^0 \\
                         x^1 \\
                         x^2 \\
                         x^3
                       \end{array}
    \right)
\ee
in case 1., while in case 2.\ the matrix
$B^\nu{}_\mu:=\nabla_\mu \Yfield ^\nu$ is nilpotent, with
$B^3=0$, so that the matrix $\Lambda_t$ associated with the
action of $\phi_t$ is
$$
\mbox{$\Lambda_t = \mathrm{Id} + t B + \frac{t^2}{2} B^2$, with } B
= \left(
\begin{array}{cccc}
                  0  & a & 0 & 0 \\
                  a & 0 & a & 0 \\
                  0 & -a & 0 & 0 \\
                  0 & 0 & 0 & 0
                \end{array}
                \right)
    \;.
$$
This shows that periodic orbits are not possible in the second
case, while in the first they are possible if and only if
$\beta_0=0$.

Smoothness of $\alpha$ can now be established by adapting the
analysis of the proof of~\cite[Proposition~3.1]{ChAPP} to the
current setting; we provide the details to exhibit some key
factorizations needed in the arguments that follow: Consider a
covering of
$$
 \mcA:=\{\Yfield =0\}
$$
by domains of definition $\mcO$ of smooth coordinate systems
$x^A$, $A=0,1$, and for $q\in \mcO$ let $x^a$, $a=2,3$, denote
normal coordinates \underline{on} $\exp_q\{(T_q\mcA)^\perp\}$.
Note that the coordinates $(x^2,x^3)$ here are \emph{not}
identical with the ones in \eq{Xfcase}-\eq{Xfcase2}, but the
$x^a|_{\exp_p\{(T_p\mcA)^\perp \}}$'s coincide, where   $p$ is
as in the analysis leading to \eq{Xfcase}-\eq{Xfcase2}. We have
just seen that $\mcA$ is a smooth timelike submanifold of
$\mcM$; and $\mcA$ is totally geodesic (in the sense of having
vanishing second fundamental form) by standard arguments. Set
$(x^\mu)=( x^A,x^a)$, and
\bel{27I.2}
 \trho = \sqrt{(x^2)^2 + (x^3)^2}
 \;.
\ee
We have  the following local form of the metric
\newcommand{\gA}{\mathring g}
\bel{afp0}
 g =  \underbrace{\gA_{AB}dx^A dx^B}_{=:\gA} + \sum_{a=2}^{3} (dx^a)^2
 +\sum_{A,a}O(\trho)dx^Adx^a + \sum_{A,B}O(\trho^2)dx^Adx^B
 +\sum_{a,b}O(\trho^2)dx^adx^b
  \;,
\ee
with $\gA$ the (Lorentzian) metric induced by $g$ on $\mcA $.
The $O(\trho^2)$ character of the $dx^a dx^b$ error terms is
standard in normal coordinates;  the $O(\trho )$ character of
the $dx^a dx^A$ error terms comes from orthogonality; the
$O(\trho^2)$ character of the $dx^A dx^B$ error terms follows
from the totally geodesic character of $\mcA $. The Killing
vector field $\Yfield $ takes the form $\Yfield  =
x^3\partial_2 - x^2\partial_3=\partial_\tvarphi$,  where
\bel{polar}
 (x^2,x^3)=(\trho \cos \tvarphi, \trho \sin
 \tvarphi)
 \;.
\ee
When expressed in terms of $\trho$ and $\tvarphi$, the
functions $g _{\mu\nu}:=g (\partial_{x^\mu},\partial_{ x^\nu})$
are smooth functions of the $x^\mu$'s. Let $R_\pi$ denote a
rotation by $\pi$ in the $(x^a)$-planes; $R_\pi$ is obtained by
flowing along $\Yfield $ a parameter-time $\pi$ and is
therefore an isometry, leading to
\begin{align*}
 g _{ab}(x^A,-x^2,-x^3)
	&= g_{ab}(x^A, x^2, x^3)
	\;,\\
 g _{AB}(x^A,-x^2,-x^3)
	&= g_{AB}(x^A, x^2, x^3)
	\;,\\
 g _{Aa}(x^A,-x^2,-x^3)
	&= -g_{Aa}(x^A, x^2, x^3)
 \;.
\end{align*}
In particular all odd-order derivatives of $g_{ab}$ with
respect to the $x^a$'s vanish at $\{x^a=0\}$, \emph{etc.} Those
symmetry properties together with Borel's summation Lemma imply
that there exist smooth fields $b_{AB}(x^C,s)$,
$\gamma_A(x^C,s)$, and $\gamma(x^C,s)$   such that
$$g _{AB}(x^C,x^2,x^3)= b_{AB}(x^C,\trho^2 )\;,$$
$$ \left(g _{Ab}\Yfield ^b\right)(x^B,x^2,x^3)= \trho ^2 \gamma_{A}(x^B,\trho^2 )\;,$$
\bel{27I.1}
 u(x^A,x^2,x^3):=\sqrt{\left(g (\Yfield ,\Yfield )\right)(x^A,x^2,x^3)}= \trho \left(1+\trho ^2 \gamma(x^A,\trho^2 )\right)
 \;.
\ee

Similarly, let $n=x^a\partial_a$, then $g _{ab}\Yfield ^a n^b $
and $g _{ab}n^a n^b$ are smooth functions invariant under the
flow of $\Yfield $, with $g _{ab}\Yfield ^a n^b=(g
_{ab}-\delta_{ab})\Yfield ^a n^b= O(\trho ^4)$, $g _{ab}n^a
n^b=\trho ^2+O(\trho ^4)$, hence there exist smooth functions
$\zeta(x^A,s)$ and $\sigma(x^A,s)$ such that
$$
 \left(g _{ab}\Yfield ^a n^b\right)(x^A,x^2,x^3)= \trho ^4 \zeta(x^A,\trho^2 )\;,
$$
$$
 \left(g _{ab}n^a n^b\right)(x^A,x^2,x^3)=\trho ^2(1 + \trho ^2
 \sigma(x^A,\trho^2 ))
 \;.
$$
We note similar formulae for the Maxwell two-form $\CF$ and its
Hodge-dual $*\CF$:
\bel{3II.1}
 \left(\CF _{ab}\Yfield ^a n^b\right)(x^A,x^2,x^3)= \trho ^2 \hat \zeta(x^A,\trho^2 )
 \;, \quad
 \left(*\CF _{ab}\Yfield ^a n^b\right)(x^A,x^2,x^3)= \trho ^2 \tilde \zeta(x^A,\trho^2 )
 \;,
\ee
\bel{3II.2}
 \left(\CF _{Ab}\Yfield ^b\right)(x^B,x^2,x^3)= \trho ^2 \hat \gamma_{A}(x^B,\trho^2 )
 \;,
 \quad
 \left(*\CF _{Ab}\Yfield ^b\right)(x^B,x^2,x^3)= \trho ^2 \tilde \gamma_{A}(x^B,\trho^2 )
 \;,
\ee
\bel{3II.3}
 \left(\CF _{Ab}n ^b\right)(x^B,x^2,x^3)= \trho ^2 \check \gamma_{A}(x^B,\trho^2 )
 \;,
 \quad
 \left(*\CF _{Ab}n ^b\right)(x^B,x^2,x^3)= \trho ^2 \dot \gamma_{A}(x^B,\trho^2 )
 \;,
\ee
for some smooth sphere functions $\hat \zeta$, $\tilde \zeta$,
$\hat \gamma_A$, $\tilde \gamma_A$, $\check \gamma_A$ and $\dot
\gamma_A$.

In the same fashion one finds existence of a smooth one-form
$\lambda_A(s,x^b)dx^A$ such that
$$
 \left(g _{Aa}n^a \right)(x^1,x^2,x^b)= \trho ^2
 \lambda_A(\trho ^2,x^b)
 \;.
$$
In polar coordinates \eq{polar} one
therefore obtains
$$
  g (\Yfield ,\cdot)= \trho ^2\left( (1+\trho ^2\,\gamma)^2 d\tvarphi + \zeta\, \trho \, d\trho  +
 \gamma_A \, dx^A\right)
 \;.
 $$
Writing $g $ in the form
\begin{equation} \label{e1.3bn}
g  = u^{2}( d\tvarphi  + \underbrace{\chi_j\,dy^j}_{=:\chi} )^{2} + \gamma_{jk}\,dy^j \,dy^k\;,
\end{equation}
with $y^j=( x^A,\trho)$, one has $ g (\Yfield ,\cdot)
=u^2(d\tvarphi +\chi_j\,dy^j)$ leading to
$$
 \chi = \frac {\trho \zeta}{ (1+\trho ^2\,\gamma)^2}\,d \trho   +
 \frac{\gamma_A}{(1+\trho ^2\psi)^{2}}\,dx^A\;,
$$
\bel{hred}\gamma_{jk}\,dy^j\, dy^k
	= (1+\trho ^2\,\sigma)  d  \trho ^2  + b_{AB}\,dx^A \,dx^B
+ 2 \lambda_A\,\trho \,d \trho\,dx^A- u^2
 \chi_i \,\chi_j\, dy^i \,dy^j
 \;,
\ee
in particular the functions $\gamma_{\trho \trho}$,
$\gamma_{AB}$, and $\gamma_{A\trho}/\trho$ are smooth functions
of $\trho ^2$ and $x^A$.

We have proved:

\begin{Proposition}
\label{Lnohg} The one-form $\chi$ defined in \eq{e1.3bn}
extends smoothly to the rotation axis $\mcA = \{\Yfield =0\}$.
\end{Proposition}

In particular
$$ \frac{g_{v \tvarphi}}{g_{\tvarphi \tvarphi}} =
 \chi(\partial_v)
$$
is a smooth function on space-time. Since it vanishes at $\tilr
=0$, the quotient $g_{v \tvarphi}/(\tilr g_{\tvarphi
\tvarphi})$ is also a smooth function on space-time by Taylor's
theorem. Hence the function $\alpha$ defined in \eq{19XII.1} is
smooth. Smoothness of $A$ and $\lambda$ as in \eq{25I.1}
follows.

In the coordinate system adapted to the horizon as in
\eq{10XII.5}, the intersection of the axis $\mcA$ and of the
Killing horizon corresponds to $\sin \ttheta=0$. To see that
this remains true in a neighbourhood of the horizon, recall
that the construction of the Gauss normal coordinates in
\eq{10XII.1} involves the family of null geodesics normal to
the section $S:=\{v=0\}$ of the connected component of the
future event horizon under consideration: the
local coordinates $(\ttheta,\tvarphi)$ on $S$ are first
Lie-propagated to $ \dot J^-(S)$ along the normal null
geodesics, and then to a neighbourhood of the Killing horizon
along the flow of $\partial_v$. Since $S$ is invariant under
the action of $\Uone$, so is its normal bundle. It follows from
\eq{29I.1} that, at the north and south poles of $S$, which are
fixed points of the rotational Killing vector $\Yfield $, those
normal geodesics are initially tangent to $\mcA$. But $\mcA$ is
totally geodesic, so in fact those geodesics remain on $\mcA$:
one of them is the generator of the event horizon, the second
one is the one which is used to propagate the coordinates
$(\ttheta,\tvarphi)$ away from the horizon. Now, $\mcA$ is also
invariant under the flow of $\partial_v$, which is tangent on
$\mcA$ to that null normal geodesic to $S$ which coincides with
the generator of the horizon. Thus $\partial_v$ is transversal
to the other null geodesic on $\mcA$, so flowing this other
geodesic along $\partial_v$ fills out a neighbourhood of this
geodesic within $\mcA$. Since $\ttheta$ is constant along the
flow of $\partial_v$, we conclude that $\sin\ttheta=0$ on
$\mcA$. Finally, e.g. by dimension considerations, we obtain
that $\{\sin\ttheta=0\}$ coincides with $\mcA=\{\trho=0\}$ in a
collar neighbourhood of $S$.

So, near $\ttheta=0$ the function $\trho$ of \eq{27I.2} is
equivalent to $\ttheta$, which is equivalent to $\sin \ttheta$,
and by the arguments above for small $\ttheta$ we have
\bel{27I.3}
   \frac {\trho}{\sin \ttheta} =  \tilde f(\tilr,
     \ttheta  )
 \;,
\ee
for some  function $\tilde f$, smooth in its arguments, bounded
away from zero, and which can be smoothly extended to an even
function of $\ttheta$ across zero. Similarly near $\ttheta=\pi$
the function $\trho$ is equivalent to $\pi- \ttheta$, which is
again equivalent to $\sin \ttheta$ near $\ttheta=\pi$, and so
the function $f$ in \eq{27I.3} extends smoothly across $\theta
= \pi$ to a function which is bounded away from zero and even
in $\pi-\theta$ for $\theta$ close to $\pi$. Since \eq{27I.3}
is trivial away from the zeros of $\sin \ttheta$, we conclude
that \eq{27I.3} holds everywhere.

Functions of $\ttheta\in[0,\pi]$ with the smooth even extension
properties near zero and $\pi$, as just described in the last
paragraph, will be called \emph{sphere functions}: indeed, a
function of $\ttheta$ defines a smooth function on a sphere if
and only if it is a sphere-function in the sense just defined.

Equations \eq{27I.1} and \eq{27I.4} lead us to
\bel{27I.4}
 \frac{h_{\tvarphi\tvarphi}}{\sin^2 \ttheta} =   f(\tilr,
    \ttheta )
 \;,
\ee
for some sphere function $f$, smooth in its arguments, and
bounded away from zero.

It also follows from what has been said so far that the
functions $\alpha$, $\lambda$, $F$,
$h_{\tvarphi\tvarphi}/\sin^2\ttheta$, $h_{\ttheta \ttheta}$,
$h_{\tvarphi \ttheta}/\sin\ttheta$, and $\lambda/\sin\ttheta$
are smooth sphere functions of $\tilr$ and $ \ttheta$.

As the next step, we modify the coordinate $\tilr $ to a new
coordinate $\hr$ by setting
\bel{10XII.2}
 d\hr = e^{\tilde \chi} (\tilr ,\ttheta)(d\tilr  + \tilr  \lambda( \tilr,\ttheta) d\ttheta)
 \;,
\ee
normalized so that
$$
 \hr(\tilr =0, \ttheta= 0)=0
	\;.
$$
Equivalently,
\bel{11XII.2}
 \partial_{\tilr} \hr = e^{\tilde \chi} \;,\quad
 \partial_{\ttheta} \hr = \tilr e^{\tilde \chi} \lambda
 \;.
\ee
The integrability conditions for $\hr$ give
\bel{11XII.1}
  \partial_{\ttheta}\tilde \chi -  \tilr\lambda \partial_{\tilr}   \tilde \chi =  \partial_{\tilr}(\tilr  \lambda)
  \;,
\ee
which can be solved by shooting characteristics from the north
pole $\ttheta=0$, where we impose $\tilde \chi =0$. Again
smoothness of $\tilde \chi$ and of $\hat r$ at the north and
south poles requires justification: Since $\lambda/\sin
\ttheta$ is a smooth sphere function of $\tilr $ and $
\ttheta$, by matching powers in a power-series expansion of
$\tilde \chi$ in \eq{11XII.1} one finds that $\tilde \chi$ is a
smooth sphere function of $\tilr $ and $  \ttheta$. In other
words, for each $\tilr$, $\tilde \chi$ defines naturally a
smooth function on $S^2$. A similar argument applies to
\eq{11XII.1}.

Since $\partial_{\ttheta}  \hr =0$ at $\tilr=0$ from
\eq{10XII.2}, we have
$$
 \hr(\tilr =0, \ttheta)=0
$$
for all $\ttheta$. Since $\tchi$ is a smooth function on
$I_{\tilr}\times S^2$, where $I_{\tilr}$ is the interval of
definition of $\tilr$,   \eq{11XII.2} implies that both
\bel{29I.3}
 \mbox{ $\displaystyle \frac{\hat r}{\tilr}$ and  $\displaystyle \frac{\tilr}{\hr}$}
\ee
are smooth functions
on $I_{\tilr}\times S^2$ near $\{\hat r =0\}$.

To summarize, we have shown:
\begin{Proposition}\label{NHMetric}
Near a spherical degenerate Killing horizon in an axially
symmetric spacetime  the metric can be written in the form
\begin{eqnarray}
 g &=& - \hat r^2\,F(\hr,{\ttheta} )\,dv^2 + 2\,\psi(\hr,{\ttheta} )\,dv\,d\hr
	+ h_{{\tvarphi}{\tvarphi}}(\hr,{\ttheta} )(d{\tvarphi} + \hr\,\alpha(\hr,{\ttheta} )\,dv)^2
		\nonumber
			\\
	&&
	+ h_{{\ttheta} {\tvarphi}}(\hr,{\ttheta} )\,(d{\tvarphi} + \hr\,\alpha(\hr,{\ttheta} )d{v} )d{\ttheta}
  + h_{{\ttheta} {\ttheta} }(\hr,{\ttheta} )\,d{\ttheta} ^2
  \;,
\label{SimplifiedMetric}
\end{eqnarray}
where $\partial_v$ is the  Killing field defining the Killing
horizon, $\partial_{\tvarphi}$ is the axial Killing field, the
horizon is at $\hr = 0$, $({\tvarphi},{\ttheta} )$ parameterize
a two-dimensional spherical cross section of the horizon, and
$F$, $\alpha$, $\psi$,
$h_{{\tvarphi}{\tvarphi}}/\sin^2\ttheta$, $h_{{\ttheta}
{\ttheta} }$, $h_{{\tvarphi}{\ttheta} }/\sin\ttheta$ (and hence
also $\det h_{ab}/\sin^2\ttheta$)
are smooth sphere functions in a neighbourhood of $\hr = 0$.

Similarly, for any anti-symmetric tensor $\CF$ the functions
$\CF_{v \tvarphi}/\sin^2\ttheta$, $\CF_{r
\tvarphi}/\sin^2\ttheta$, $\CF_{v \ttheta}/\sin \ttheta$,
$\CF_{\ttheta\tvarphi}/\sin \ttheta$ and   $\CF_{v \hr}$ are
smooth sphere functions.
\end{Proposition}

\section{Geometric analysis near a degenerate horizon}
 \label{Sgeomana}

In this section, we would like to extract geometric information
near a degenerate horizon $\mcE_0$ in an axially symmetric and
stationary electrovacuum space-time $(\mcM,g)$ using the metric
form \eqref{SimplifiedMetric}.

More precisely, it has been shown in~\cite{Hajicek3Remarks} in
vacuum, and in~\cite{LP1} in electrovacuum, that the
near-horizon geometry  is determined uniquely by the area $A_0$
of a cross section $S_0$ of the horizon $\mcE_0$, the electric
charge $q_e$ and the magnetic charge $q_b$ of the horizon. For
convenience of notation we introduce the \emph{area radius} of
the horizon:
\begin{equation}
r_0 = \sqrt{\frac{A_0}{4\pi}}
	\;.
\label{r0}
\end{equation}
Note that, by the near-horizon analysis in~\cite{LP1},
\begin{equation}
r_0^2 \geq q_e^2 + q_b^2
	\;.
\label{ValidRange}
\end{equation}
%

\subsection{The near-horizon limit in vacuum} \label{ssKerr}

Assume that $(\mcM,g)$ is a vacuum space-time, so $q_e = q_b =
0$. The near-horizon geometry of the extreme Kerr solution
which has horizon area $A_0$ is given by (see, e.g.,
\cite{BardeenHorowitz}):
\[
g_{\mbox{\scriptsize NHK}} = \frac{1 + \cos^2\theta}{2}\Big[-\frac{\hr^2}{r_0^2}\,dt^2 + \frac{r_0^2}{\hr^2}\,d\hr^2 + r_0^2\,d\theta^2\Big] + \frac{2r_0^2\,\sin^2\theta}{1 + \cos^2\theta} \Big(d\phi + \frac{\hr}{r_0^2}dt\Big)^2
	\;,
\]
where $(t,\hr + r_0/\sqrt{2},\theta,\phi)$ is the Boyer-Lindquist coordinate system for the Kerr solution.

By the change of variables
\[
v = t - \frac{r_0^2}{\hr}
	\;,\qquad \tvarphi = \phi - \log \left(\frac{\hr}{r_0}\right)
 \;,
\]
the above metric can be rewritten as
\[
g_{\mbox{\scriptsize NHK}} = \frac{1 + \cos^2\theta}{2}\Big[-\frac{\hr^2}{r_0^2}\,dv^2 + 2\,dv\,d\hr + r_0^2\,d\theta^2\Big] + \frac{2r_0^2\,\sin^2\theta}{1 + \cos^2\theta} \Big(d\tvarphi + \frac{\hr}{r_0^2}dv\Big)^2
	\;.
\]

We then use the results in~\cite{Hajicek3Remarks} and the
analysis in Section \ref{SnhsXII.1} to obtain, in a
neighbourhood of $\mcE_0$, a null Gaussian coordinate system
$(v,\varphi,\hr,\ttheta)$ such that the metric $g$ takes the
form \eqref{SimplifiedMetric} and the coordinates $(\ttheta,
\varphi)$ agree with the coordinate $(\theta,\varphi)$ of the
above Kerr metric at $\hr=0$. Furthermore,
\begin{align}
F(0,\ttheta)
	&= \frac{1}{2r_0^2}\,(1 + \cos^2 \ttheta)  \;,\label{r=0-A}\\
\psi(0,\ttheta)
	&= \frac{1}{2}\,(1 + \cos^2 \ttheta) \;,\label{r=0-psi}\\
h_{\tvarphi\tvarphi}(0,\ttheta)
	&= \frac{2r_0^2\sin^2\ttheta}{1 + \cos^2\ttheta} \;,\label{r=0-hxx}\\
h_{\tvarphi\ttheta}(0,\ttheta)
	&= 0
 \;,
 \label{r=0-hxy2}
\\
 h_{\ttheta\ttheta}(0,\ttheta)
	&= \frac{1}{2}\,r_0^2\,(1 + \cos^2\ttheta)  \;,\label{r=0-hyy}
\\
 \alpha(0,\ttheta)
	&= \frac{1}{r_0^2}
 \;.
 \label{r=0-alpha}
\end{align}

Observe that equations \eqref{r=0-hxx}-\eqref{r=0-hyy} together
with Proposition \ref{NHMetric} allow us to write
\begin{equation}
 h_{\tvarphi\tvarphi} = \frac{2r_0^2\sin^2\ttheta}{1 + \cos^2\ttheta}\,\beta_{\varphi\varphi}
	\;, \qquad \det h = r_0^4\,\sin^2\ttheta\,\beta
	\;,
\label{KeyFtrztn}
\end{equation}
for some smooth sphere functions $\beta_{\tvarphi\tvarphi}$ and
$\beta$ of $(\hr, \ttheta)$, which satisfy
$\beta_{\tvarphi\tvarphi}(0,\ttheta) \equiv \beta(0, \ttheta)
\equiv 1$.

\subsection{The near-horizon limit in electrovacuum}
 \label{ssKerrNewmanme}

In the general case where $(\mcM,g)$ is electrovacuum,
by~\cite{LP1}, the near-horizon geometry is characterized by
that of the Kerr-Newman solution which has the same horizon
area parameter $A_0$ and charge parameters $q_e$ and $q_b$. In
the Kerr-Newman  case, the near-horizon fields can be obtained
by first applying a duality rotation to $\CF_{\mbox{\scriptsize
KN}}$ (to account for the magnetic charge), and then
calculating the near-horizon limit. Using, e.g.,
\cite[pp.79-80]{Heusler:book} one finds
(compare~\cite{BardeenHorowitz}):
\begin{align*}
g_{\mbox{\scriptsize NHKN}}
	&= \frac{\mzero ^2 + \azero^2\,\cos^2\theta}{r_0^2}\,\Big[-\frac{\hr^2}{r_0^2}\,dt^2 + \frac{r_0^2}{\hr^2}\,d\hr^2 + r_0^2\,d\theta^2\Big]\\
		&\qquad\qquad + \frac{r_0^4\,\sin^2\theta}{\mzero ^2 + \azero^2\,\cos^2\theta}\,\Big(d\phi
 + \frac{2\azero \,\mzero \,\hr}{r_0^4}\,dt\Big)^2
	\;,\\
\CF_{\mbox{\scriptsize NHKN}}
	&= q_e\Big\{-\frac{2\,\azero\,\mzero\,r_0^2\,\sin\theta\,\cos\theta}{(\mzero^2 + \azero^2\,\cos^2\theta)^2}\,d\phi \wedge d\theta + \frac{(\mzero^2 - \azero^2\,\cos^2\theta)}{r_0^2(\mzero^2 + \azero^2\,\cos^2\theta)}\,d\hr \wedge dt
\\
		&\qquad\qquad + \frac{4\,\azero^2\,\mzero^2\,\hr\,\sin\theta\,\cos\theta}{r_0^2\,(\mzero^2 + \azero^2\,\cos^2\theta)^2}\,d\theta \wedge dt\Big\}
\\
		&+ q_b\Big\{\frac{r_0^2(\mzero^2 - \azero^2\,\cos^2\theta)\sin\theta}{(\mzero^2 + \azero^2\,\cos^2\theta)^2}\, d\phi \wedge d\theta + \frac{2\azero\,\mzero\,\cos\theta}{r_0^2(\mzero^2 + \azero^2\,\cos^2\theta)}\,d\hr \wedge dt
\\
		&\qquad\qquad - \frac{2\,\azero\,\mzero\,(\mzero^2 - \azero^2\,\cos^2\theta)\,\hr\,\sin\theta}{r_0^2(\mzero^2 + \azero^2\,\cos^2\theta)^2}\,d\theta \wedge dt\Big\}
	\;.
\end{align*}
Here $r_0$ is as in \eq{r0}, $\azero = \sqrt{(r_0^2 - q_b^2 -
q_b^2)/2}$ and $\mzero =\sqrt{\azero ^2 + q_e^2+ q_b^2}$. Note
that the sign of $\azero$ is not determined in~\cite{LP1}, but
we can always make it positive using the transformation $\phi
\mapsto -\phi$.

Introducing the change of variables
\[
v = t - \frac{r_0^2}{\hr}
	\;,\qquad \tvarphi = \phi - \frac{2\azero \,\mzero }{r_0^2}\,\log \hr,
\]
we obtain
\begin{align}
 \nonumber
g_{\mbox{\scriptsize NHKN}}
	&= \frac{\mzero^2
 + \azero^2\,\cos^2\theta}{r_0^2}\Big[-\frac{\hr^2}{r_0^2}\,dv^2 + 2dv\,d\hr^2 + r_0^2\,d\theta^2\Big]\\
		&\qquad\qquad + \frac{r_0^4\,\sin^2\theta}{\mzero^2
 + \azero^2\,\cos^2\theta}\,\Big(d\tvarphi + \frac{2\azero\,\mzero\,\hr}{r_0^4}\,dv\Big)^2
	\;,
 \label{6II.1}
\\
\CF_{\mbox{\scriptsize NHKN}}
	&= q_e\Big\{-\frac{2\,\azero\,\mzero\,r_0^2\,\sin\theta\,\cos\theta}{(\mzero^2 + \azero^2\,\cos^2\theta)^2}\,d\tvarphi \wedge d\theta + \frac{(\mzero^2 - \azero^2\,\cos^2\theta)}{r_0^2(\mzero^2 + \azero^2\,\cos^2\theta)}\,d\hr \wedge dv	\nonumber\\
		&\qquad\qquad + \frac{4\,\azero^2\,\mzero^2\,\hr\,\sin\theta\,\cos\theta}{r_0^2\,(\mzero^2 + \azero^2\,\cos^2\theta)^2}\,d\theta \wedge dv\Big\}	\nonumber\\
		&\qquad\qquad + q_b\Big\{\frac{r_0^2(\mzero^2 - \azero^2\,\cos^2\theta)\sin\theta}{(\mzero^2 + \azero^2\,\cos^2\theta)^2}\, d\tvarphi \wedge d\theta + \frac{2\azero\,\mzero\,\cos\theta}{r_0^2(\mzero^2 + \azero^2\,\cos^2\theta)}\,d\hr \wedge dv	\nonumber\\
		&\qquad\qquad - \frac{2\,\azero\,\mzero\,(\mzero^2 - \azero^2\,\cos^2\theta)\,\hr\,\sin\theta}{r_0^2(\mzero^2 + \azero^2\,\cos^2\theta)^2}\,d\theta \wedge dv\Big\}
	\;.
\end{align}

Thus, as already explained, we can select a null Gaussian
coordinate system $(v,\hr,\ttheta,\tvarphi)$ in a neighbourhood
of $\mcE_0$ in $\mcM$ such that $g$ takes the form
\eqref{SimplifiedMetric} there, the coordinates
$(\ttheta,\tvarphi)$ coincide  with the coordinates
$(\theta,\varphi)$ as in \eq{6II.1} on $\mcE_0$, and
\begin{align}
F(0,\ttheta)
	&= \frac{\mzero^2 + \azero^2\,\cos^2\ttheta}{r_0^4}  \;,\label{r=0-A-ev}\\
\psi(0,\ttheta)
	&= \frac{\mzero^2 + \azero^2\,\cos^2\ttheta}{r_0^2} \;,\label{r=0-psi-ev}\\
h_{\tvarphi\tvarphi}(0,\ttheta)
	&= \frac{r_0^4\sin^2\ttheta}{\mzero^2 + \azero^2\,\cos^2\ttheta} \;,\label{r=0-hxx-ev}\\
h_{\tvarphi\ttheta}(0,\ttheta)
	&= 0
 \;,
 \label{r=0-hxy}
\\
 h_{\ttheta\ttheta}(0,\ttheta)
	&= \mzero^2 + \azero^2\,\cos^2\ttheta  \;,\label{r=0-hyy-ev}
\\
 \alpha(0,\ttheta)
	&= \frac{2\azero\,\mzero}{r_0^4}
 \;.
 \label{r=0-alpha-ev}
\end{align}

Moreover, by Proposition \ref{NHMetric} we have
\begin{equation}
 h_{\tvarphi\tvarphi} = \frac{r_0^4\sin^2\ttheta}{\mzero^2 + \azero^2\,\cos^2\ttheta}\,\beta_{\tvarphi\tvarphi}
	\;,\qquad  \det h = r_0^4\sin^2\ttheta\,\beta
	\;,
\label{KeyFtrztn-ev}
\end{equation}
for some smooth sphere functions $\beta_{\tvarphi\tvarphi}$ and
$\beta$ of $(\hr, \ttheta)$ which satisfy
$\beta_{\tvarphi\tvarphi}(0, \ttheta) \equiv \beta(0, \ttheta)
\equiv 1$.

\subsection{The orbit-space metric}\label{SSorbit}

In the following, we use $x^A$ as the dummy variable for $r$
and ${\ttheta} $ and $x^a$  as the dummy variable for $v$ and
${\tvarphi}$.  This should \emph{not} be confused with the
coordinates $(x^A,x^a)$ of the proof of
Proposition~\ref{Lnohg}.

The Killing part of the metric $g$ is defined as
\begin{equation}
g_\parallel = - \hr^2\,F(\hr,{\ttheta} )\,dv^2 + h_{{\tvarphi}{\tvarphi}}(\hr,{\ttheta} )(d{\tvarphi} + \hr\,\alpha(r)\,dv)^2.
\label{KillingPart}
\end{equation}
Note that
\[
\det g_\parallel = - \hat r^2\,F\,h_{{\tvarphi}{\tvarphi}}.
\]
In particular, $g_\parallel$ is Lorentzian for $\hr\ne 0$ if
and only if $ Ah_{{\tvarphi}{\tvarphi}}$ is non-negative. In
electrovacuum this follows from  \eq{27I.4} and from  the
analysis of the near-horizon geometry
in~\cite{Hajicek3Remarks,LP1} (compare \eq{r=0-A}  and
\eqref{r=0-A-ev}). Alternatively, one can simply assume that
this is true and carry-on the analysis from there.

The orbit-space metric
$q$ is defined as
\[
q_{AB} = g_{AB} - g_{\parallel}^{ab}\,g_{Aa}\,g_{Bb} \;,
\]
where $ g_{\parallel}^{ab}$ is the matrix inverse to
$g_{\parallel}(\partial_a,\partial_b)$. In matrix notation
\eqref{KillingPart} reads
\[
g_{\parallel} = \matrixform{cc}{
\hr^2(-F + \alpha^2\,h_{{\tvarphi}{\tvarphi}}) & \hr\,\alpha\,h_{{\tvarphi}{\tvarphi}}\\
\hr\,\alpha\,h_{{\tvarphi}{\tvarphi}} & h_{{\tvarphi}{\tvarphi}}
}
	\;,
\]
and so its inverse reads
\[
g_{\parallel}^{-1} = -\frac{1}{\hr^2\,F\,h_{{\tvarphi}{\tvarphi}}}\matrixform{cc}{
h_{{\tvarphi}{\tvarphi}} & -\hr\,\alpha\,h_{{\tvarphi}{\tvarphi}}\\
-\hr\,\alpha\,h_{{\tvarphi}{\tvarphi}} & \hr^2(-F + \alpha^2\,h_{{\tvarphi}{\tvarphi}})
}
	\;.
\]
The orbit-space metric is then
\begin{equation}
q = q_{\hr\hr}\,d\hr ^2 + q_{{\ttheta} {\ttheta} }\,d{\ttheta} ^2 =
 F^{-1}\,\psi^2\,\frac{d\hr ^2}{\hr^2} +
 \frac{\det h}{h_{{\tvarphi}{\tvarphi}}}
 \,d{\ttheta} ^2       \;.
\label{QuotientMetric}
\end{equation}

By \eqref{r=0-A-ev}-\eqref{r=0-alpha-ev}
we have
\begin{equation}
q_{\hr\hr} = \frac{1}{\hr^2}\Big( \mzero^2 + \azero^2\cos^2{\ttheta}   + O(\hr)\Big)
 \;,
 \qquad q_{{\ttheta} {\ttheta} } = \mzero^2 + \azero^2\cos^2{\ttheta}  + O(\hr)
  \;,
\label{QMetric-Asymp}
\end{equation}
with the error terms meant for small $\hr$.

Now, define the function $\rho $ by
\begin{equation}
\rho = \sqrt{- \det g_\parallel} = \hr\,\sqrt{F\,h_{{\tvarphi}{\tvarphi}}}      \;.
\label{rho-Def}
\end{equation}
By \eqref{r=0-A-ev} and \eqref{KeyFtrztn-ev}  we have
\begin{equation}
\rho = \hr\,\tilde\beta_\rho\,\sin{\ttheta}
\label{rho-Asymp}
\end{equation}
for some smooth sphere function $\tilde\beta_\rho =
\tilde\beta_\rho(\hr ,\ttheta)$  such that
$$\tilde\beta_\rho (\hr,\ttheta) = 1+ O(\hr)
 \;.
$$
By the Einstein-Maxwell electrovacuum equation, $\rho$ is
harmonic with respect to $q$  (see, e.g.,
\cite[Section~2]{Weinstein3}). Let $z$ be minus the harmonic
conjugate of $\rho$, i.e.\ $z$ is defined up to a constant by
\beaa
  &
  \displaystyle
  z_{,\hr}
  = q^{\ttheta\ttheta}\,\sqrt{\det q}\,\rho_{,\ttheta}
 = \sqrt{\frac{h_{\tvarphi\tvarphi}}{F \det h}} \frac\psi {\hr} \,\rho_{,{\ttheta} }
 \;,
 &
\\
 \qquad
&
  \displaystyle
 z_{,{\ttheta} }
  = - q^{\hr\hr}\,\sqrt{\det q}\,\rho_{,{\hr} }
  =  - \sqrt{\frac{F \det h}{h_{\tvarphi\tvarphi}}} \frac {\hr} \psi \,\rho_{,\hr} \;.
  &
\eeaa
By \eqref{r=0-A-ev}-\eqref{r=0-hyy-ev} and \eqref{rho-Asymp},
we have
\[
  z_{,\hr}  = \gamma\,\tilde\beta_\rho\,\cos\ttheta
	\;,
 \qquad
 z_{,{\ttheta} }  =  - \frac{1}{\gamma}\,\hat r\,(\hr\,\tilde\beta_\rho)_{,\hr}\,\sin\ttheta
	\;,
\]
where $\gamma$ is some smooth positive function of
$(\hr,\ttheta)$ such that $\gamma(0,\ttheta) \equiv 1$. Thus, up to a shift by
a constant,
\begin{equation}
 z = \hr\,\tilde\beta_z\,\cos{\ttheta}
 \;,
\label{z-Asymp}
\end{equation}
where
\[
\tilde\beta_z(\hr,\ttheta) = \frac{1}{\hr}\int_0^\hr \gamma(s,\ttheta)\,\tilde\beta_{\rho}(s,\ttheta)\,ds
 = 1 + O(\hr)
 \;.
\]
Altogether, \eqref{rho-Asymp} and \eqref{z-Asymp} imply that
\[
r^2:=\rho^2 + z^2 = \hr^2 + O(\hr^3) \text{ as } \hr \rightarrow 0.
\]
We have thus proved:

\begin{Proposition}
 \label{Pkeydeg}
In $I^+$--regular, axisymmetric, stationary and electrovacuum
space-times, every degenerate component of the event horizon
corresponds to a point lying on the axis $\rho=0$ in the
$(\rho,z)$ plane.
\end{Proposition}

\begin{Remark}
 \label{RPkeydeg}
For the sake of simplicity we have stated the result under the
hypotheses of Theorem~\ref{Tmain30I.1}. However, the analysis
above only uses the following: the horizon is degenerate, and
has a spherical cross-section $S$ on which the Killing vector
$\partial_v$ has no zeros; the Killing vector field
$\partial_\varphi$ is spacelike wherever non-zero; the function
$\rho$ is harmonic with respect to the orbit-space metric $q$;
and finally
$$
 \lim_{\hr \to 0}  {\frac{F   \det h  }{ \psi^2 h_{\tvarphi\tvarphi}}}
 = 1 = \lim_{\hr \to 0} \frac{F\,h_{{\tvarphi}{\tvarphi}}}{\sin^2\ttheta}
  \;.
$$
\end{Remark}

\subsection{Global isothermal coordinates}
 \label{ssGic}

We wish to show that the functions $\rho$ and $z$ provide
global coordinates on the quotient manifold
$\doc/(\R\times\Uone)$,
 where $\doc$  is the domain of outer
communications in $(\mcM,g)$. For this we adopt the strategy
in~\cite{ChCo}, which in turn draws on~\cite{ChUone}; the
arguments there need to be extended in a non-trivial way to
cover the current setting.

Let $\mcB$ be the manifold obtained from the orbit space
$\doc/(\R\times\Uone)$ by doubling along the axis, as
in~\cite{ChUone}. The metric $q$ extends smoothly to a smooth
metric on $\mcB$, which we will also denote by $q$. In this
section, we show that the functions $\rho$ and $z$, defined in
the previous section and appropriately extended to the double,
provide \textit{global} isothermal coordinates for $(\mcB, q)$
and hence, by restriction, for $\doc/(\R\times\Uone)$.

As shown in Proposition~\ref{Pkeydeg}, in the connected case
the horizon corresponds to a point $p$  in a one-point
completion $\overline{\mcB}:=\mcB\cup \{p\}$ of $\mcB$. The
point $p$ will be denoted by $0$; the reason for this slight
abuse of notation will be clear momentarily. For configurations
with $N_d$ degenerate components of the horizon and $N_r$
non-degenerate ones, each degenerate horizon will correspond to
a point $p_i$ in a completion
$$
 \overline{\mcB} := \mcB\cup_{i=1}^{N_d}\{p_i\}\cup_{a=1}^{N_r}D_{a}^1
$$
where the $D_a^1$'s are disks corresponding to smooth boundary
components for $\overline{\mcB}$; see~\cite{ChCo} for a
detailed description of the non-degenerate components of the
event horizon. It should be noted that the point $0$ in the
former case, and the $p_i$'s in the latter case, are genuinely
\emph{not} points in $\mcB$.

In  a $\overline\mcB$--neighbourhood of each $p_i$    we
parameterize $\mcB$ by a small punctured disc
$D_{4\epsilon}\setminus\{0\} \subset \RR^2$ via the polar map
$(\hx,\hy) \mapsto (\hr,\ttheta)$, with $\hr \in(
0,4\epsilon)$. By \eqref{QMetric-Asymp}  in this region, $q$ is
conformal to
\[
 \hat q := d\hr^2 + r^2\,f(\hr,\ttheta)\,d\ttheta^2
 \;,
\]
where $f$ is a smooth sphere function such that $f(0,\ttheta)
\equiv 1$. This can be rewritten as
\beaa
 \hat q &= &  d\hr^2 +  f(\hr,\ttheta)\,(d\hx^2 + d\hy^2 - d\hr^2)
\\
 & = &
 f(\hr,\ttheta)\,(d\hx^2 + d\hy^2) +  \frac{f(\hr,\ttheta)-1}{\hr
 ^2}(\hx \,d\hx + \hy\, d\hy)^2
  \;.
\eeaa
So $\hat q$ will extend smoothly across $\hx = \hy =0$ if and
only if $f-1$ equals $\hr^2$ times a smooth function of $\hx$
and $\hy$. If this happens to be the case, we can apply
\cite[Theorem 2.9]{ChUone} to reach the desired conclusion,
Theorem~\ref{T30I.1} below. However, it is not clear that   $f$
will take this form in general, so the above strategy needs to
be revised to allow general metrics $\hat q$ as above. For this
we need to provide first some preliminary analysis.

Let $R_{\hat q}$ denote the Gaussian curvature of $\hat q$. It
is evident that $R_{\hat q}$ is smooth in all sufficiently
small punctured discs $D_{4\epsilon}\setminus \{0\}$,
$0<\epsilon<\epsilon_0$ for some $\epsilon_0$, with
\begin{equation}
 R_{\hat q} = O(\hr^{-1}) \text{ and } |DR_{\hat q}| = O(r^{-2}) \text{ for small } \hr > 0
 \;.
\label{LipDec}
\end{equation}
Moreover,   the usual formula for the scalar curvature in a
frame formalism in dimension two shows that there are functions
\begin{equation}
\mbox{$\hat f_{x}$, $\hat f_{y} \in$
 $C^\infty(D_{4\epsilon}\setminus\{0\}) \cap
 L^\infty(D_{4\epsilon})$ such that  $
 R_{\hat q} = \partial_{x} \hat f_{x} + \partial_{y} \hat f_{y}
 $.}
\label{NegSob}
\end{equation}
In particular, $R_{\hat q} \in H^{-1}(D_{4\epsilon})$. Let
$\hat u \in H^1_0(D_{4\epsilon})$ be the solution to (see e.g.
\cite[Theorem 8.3]{GT})
\[
\left\{\begin{array}{ll}
-\Delta_{\hat q} \hat u = \frac{R_{\hat q}}{2} \text{ in } D_{4\epsilon}	\;,\\
\hat u = 0 \text{ on } \partial D_{4\epsilon}	\;,
\end{array}\right.
\]
so that the metric $e^{-2\hat u}\hat q$ is flat in
$D_{2\epsilon} \setminus \{0\}$. By \eqref{NegSob} and standard
elliptic estimates (see e.g. \cite[Theorem 8.24]{GT}),
$\hat u$ is smooth in $D_{4\epsilon}\setminus\{0\}$,
$\mu$-H\"{o}lder continuous in $D_{4\epsilon}$ for some $\mu
\in (0,1)$ with
\[
 \|\hat u\|_{C^\mu(D_{2\epsilon})}
 \leq C(\epsilon,\|  f \|_{L^\infty(D_{4\epsilon})},\|\hat f_x\|_{L^\infty(D_{4\epsilon})}, \|\hat f_y\|_{L^\infty(D_{4\epsilon})})
 \;.
\]

Now, pick any $\hat u_* \in C_c^\infty(\mcB) \cap
C^\mu(\overline{\mcB})$ such that $\hat u_* \equiv u$ in the
region which is parameterized by $D_\epsilon$. Define $\tilde q
= e^{-2\hat u_*}\,\hat q$. It is readily seen that $\tilde q$
is flat near $0$. Since $\hat u$ is continuous, $\tilde q$ has
no conical singularity at the origin, and so the metric $\tilde
q$ is smooth across $0$ in an appropriate differentiable
structure (which might, or might not, coincide with the one
defined by the coordinates $\hx$ and $\hy$, but this turns out
to be irrelevant for what follows). We can now
apply~\cite[Theorem 2.19]{ChUone} to find a function $\tilde u
\in C^\infty(\overline{\mcB})$ such that $\overline
q:=e^{-2\tilde u}\,\tilde q$ is a smooth flat metric on the
complete simply connected manifold $\overline{\mcB}$. Since the
relevant equations are conformally invariant, one can ignore
the possible singularities at the $p_i$'s of the conformal
factor relating $q$ and $\overline q$, and proceed as
in~\cite{ChCo} (see in particular the argument leading from
Equation~(6.8) to Equation~(6.11) there) to show that $\rho$
and $z$ provide a global coordinate system on
$\doc/(\R\times\Uone)$. We conclude that

\begin{Theorem}
 \label{T30I.1}
 Under the hypotheses of Theorem~\ref{Tmain30I.1}, the
 area function $\rho$ and  its harmonic conjugate $-z$ form
 a global manifestly asymptotically flat coordinate system on
 $\doc/(\R\times\Uone)$.
\end{Theorem}

For the sake of completeness, we note some regularity
properties of $\hat  u$. Fix some point $p \in
D_\epsilon\setminus\{0\}$. Applying \cite[Theorem 8.32]{GT} to
$u - u(0)$ in $D_{|p|/2}(p)$ and recalling \eqref{LipDec}, we
have
\begin{align*}
|D \hat u(p)|
	&\leq C\Big[|p|^{-1}\,\|\hat u - u(0)\|_{L^\infty(D_{|p|/2}(p))} + |p|\,\|R_{\hat q}\|_{L^\infty(D_{|p|/2}(p))}\Big]\\
	&\leq C\Big[|p|^{\mu-1} + 1\Big].
\end{align*}
It follows that
\begin{equation}
 |D\hat u(p)| \leq |p|^{\mu - 1} \text{ for any } p \in D_\epsilon\setminus \{0\}
 \;.
\label{GradEst}
\end{equation}
Similarly, applying~\cite[Theorem 6.2]{GT} to $u - u(0)$ in
$D_{|p|/2}(p)$ and noting that, by \eqref{LipDec}
\[
\|R_{\hat q}\|_{C^{\mu'}(D_{|p|/2}(|p|))} \leq C\,|p|^{-1-\mu'} \text{ for any } \mu' \in (0,1],
\]
we get
\begin{align*}
|D^2 \hat u(p)|
	&\leq C\Big[|p|^{-2}\,\|\hat u - u(0)\|_{L^\infty(D_{|p|/2}(p))} + |p|^{\mu'}\,\|R_{\hat q}\|_{C^{\mu'}(D_{|p|/2}(p))}\Big]\\
	&\leq C\Big[|p|^{\mu-2} + |p|^{\mu'-1}\Big].
\end{align*}
We thus have
\begin{equation}
|D^2 \hat u(p)| \leq  |p|^{\mu - 2} \text{ for any } p \in D_\epsilon\setminus \{0\}.
\label{HessEst}
\end{equation}
%

\subsection{Hypersurface-orthogonality}

Recall that $\kfield = \partial_v$, $\Yfield  =
\partial_{\tvarphi}$. It is well-known that, in electrovacuum (see,
e.g., \cite{KundtTrumper}), the plane distribution $(\Span\{
\kfield,\Yfield\})^\perp$ is integrable; equivalently
\begin{equation}
d\kfield \wedge \kfield \wedge \Yfield = d\Yfield \wedge \kfield \wedge \Yfield = 0
	\;.
\label{HO}
\end{equation}
By direct computations, we find
\begin{align*}
d\kfield \wedge \kfield \wedge \Yfield
	&=  \Big\{\hr^2\,h_{{\tvarphi}{\tvarphi}}\Big[-\alpha\,h_{{\tvarphi}{\tvarphi}}\,[\alpha\,\psi]_{,{\ttheta} }	+ 	F_{,{\ttheta} }\,\psi - \psi_{,{\ttheta} }\,F\Big]\\
		&\qquad\qquad + \hr^3\,\alpha\,F\Big[- h_{{\tvarphi}{\tvarphi},\hr}\,h_{{\tvarphi}{\ttheta} }	+ 	h_{{\tvarphi}{\tvarphi}}\,h_{{\tvarphi}{\ttheta} ,\hr}\Big] \Big\}
	\  dv \wedge d\hr  \wedge d{\tvarphi} \wedge d{\ttheta}
 \;,
\\
 d\Yfield \wedge \kfield \wedge \Yfield
	&= \Big\{- \hr\,\psi\,h_{{\tvarphi}{\tvarphi}}^2\,\alpha_{,{\ttheta} }\\
			&\qquad\qquad +	\hr^2\,F\Big[ - h_{{\tvarphi}{\tvarphi},\hr}\,h_{{\tvarphi}{\ttheta} } + h_{{\tvarphi}{\tvarphi}}\,h_{{\tvarphi}{\ttheta} ,\hr}\Big]	\Big\}
 \ dv \wedge d\hr  \wedge d{\tvarphi} \wedge d{\ttheta}
 	\;.
\end{align*}
Thus the hypersurface orthogonality condition \eqref{HO} reads
\begin{align}
&- \psi\,h_{{\tvarphi}{\tvarphi}}^2\,\alpha_{,{\ttheta} }	+	\hr\,F\Big[ - h_{{\tvarphi}{\tvarphi},\hr}\,h_{{\tvarphi}{\ttheta} } + h_{{\tvarphi}{\tvarphi}}\,h_{{\tvarphi}{\ttheta} ,\hr}\Big]	=	0 	\;,\label{OC1}\\
&F_{,\ttheta}\,\psi + (-F + \alpha^2\,h_{\tvarphi\tvarphi})\,\psi_{,\ttheta}	=	0	\;.\label{OC2}
\end{align}

\subsection{The Ernst potential of $\partial_{\tvarphi}$ in vacuum}
 \label{ssEptv}

We now turn our attention to the second missing ingredient
required for the uniqueness argument. Namely, we will show
that, in a neighbourhood of the horizon $\mcE_0$, the harmonic
map associated to $(\mcM,g)$ lies a finite distance from that
associated to the Kerr-Newman solution which has the same
parameters $A_0$, $q_e$ and $q_b$. We start with the  special
case where $(\mcM,g)$ is vacuum. The electrovacuum case will be
considered in Section \ref{ssEptev}.

The (complex) Ernst potential associated with the
Killing vector $\Yfield =\partial_{\tvarphi}$ is defined as $X
+ i\,Y$ where
\begin{align*}
X
	 = g(\Yfield,\Yfield)    \;, \qquad
dY
	 = *(\Yfield \wedge d\Yfield)     \;,
\end{align*}
where $*$ is the Hodge operator of $g$. Here, by a common abuse
of notation, we use  the same symbol $\Yfield$ for the vector
$\Yfield$ and its metric dual $g(\Yfield, \cdot)$. The
existence of the twist potential $Y$ is a consequence of the
Einstein vacuum equations; see, e.g.,
\cite[Section~2]{Weinstein1}.

The reference Kerr metric has been chosen to have the same area
radius $r_0$ as the metric under consideration, and so from
\eqref{KeyFtrztn}, we have
\begin{equation}
X = \frac{2r_0^2\,\sin^2\ttheta}{1 + \cos^2\ttheta}(1 + O(\hr)).
\label{X-Ftrztn}
\end{equation}

To obtain the twist potential $Y$, a computation gives
\begin{align*}
\Yfield \wedge d\Yfield
	&= -h_{{\tvarphi}{\tvarphi}}^2\,[\hr\,\alpha]_{,\hr}\,dv  \wedge d\hr  \wedge d{\tvarphi}     +      \hr\,h_{{\tvarphi}{\tvarphi}}^2\,\alpha_{,{\ttheta} }\,dv  \wedge d{\tvarphi} \wedge d{\ttheta} \\
		&\qquad\qquad + \Big[\hr\,\alpha\,h_{{\tvarphi}{\tvarphi}}\,h_{{\tvarphi}{\ttheta} ,\hr} - h_{{\tvarphi}{\ttheta} }\,[\hr\,\alpha\,h_{{\tvarphi}{\tvarphi}}]_{,\hr}\Big]\,dv  \wedge d\hr  \wedge d{\ttheta} \\
		&\qquad\qquad + \Big[ - h_{{\tvarphi}{\tvarphi}}\,h_{{\tvarphi}{\ttheta} ,\hr} + h_{{\tvarphi}{\ttheta} }\,h_{{\tvarphi}{\tvarphi},\hr}\Big]\,d\hr   \wedge d{\tvarphi} \wedge d{\ttheta}
 \;.
\end{align*}
Using
$$
 \det g_{\mu\nu} = -\psi^2 \det h
 \;,
$$
we are led to
\begin{align*}
*(\Yfield \wedge d\Yfield)
	&= \frac{dv}{\sqrt{\det h}}\Big\{- \hr\,h_{{\tvarphi}{\tvarphi}}^2\,\alpha_{,{\ttheta} } + \hr^2\,F\,\psi^{-1}\,\Big[-h_{{\tvarphi}{\tvarphi},\hr}\,h_{{\tvarphi}{\ttheta} } + h_{{\tvarphi}{\tvarphi}}\,h_{{\tvarphi}{\ttheta} ,\hr}\Big]	\Big\}\\
	&\qquad - \frac{d\hr }{\sqrt{\det h}}\Big[ - h_{{\tvarphi}{\tvarphi},\hr}\,h_{{\tvarphi}{\ttheta} } + h_{{\tvarphi}{\tvarphi}}\,h_{{\tvarphi}{\ttheta} ,\hr}\Big]\\
	&\qquad + d{\ttheta} \,h_{{\tvarphi}{\tvarphi}}\,\sqrt{\det h}\,\psi^{-1}[\hr\,\alpha]_{,\hr}	\;.
\end{align*}
In the above formula, the $dv$ component must vanish. This is a
consequence of one of the hypersurface orthogonality
conditions, namely that $d\Yfield \wedge \kfield \wedge \Yfield
= 0$ (see \eqref{OC1}). Thus,
\begin{multline}
*(\Yfield \wedge d\Yfield)
	= - \frac{1}{\sqrt{\det h}}\Big[ - h_{{\tvarphi}{\tvarphi},\hr}\,h_{{\tvarphi}{\ttheta} } + h_{{\tvarphi}{\tvarphi}}\,h_{{\tvarphi}{\ttheta} ,\hr}\Big]\,d\hr\\
		+	 h_{{\tvarphi}{\tvarphi}}\sqrt{\det h}\,\psi^{-1}[\hr\,\alpha]_{,\hr}\,d{\ttheta}	\;.
\label{twist}
\end{multline}
Now, as $dY = *(\Yfield \wedge d\Yfield)$, the relations
\eqref{r=0-psi}, \eqref{r=0-alpha} and \eqref{KeyFtrztn} imply
\begin{equation}
Y_{,\hr} = \gamma_{\hr}\,\sin\ttheta
	\;, \qquad
 Y_{,\ttheta} = 4r_0^2\,\gamma_{\ttheta}\,\frac{\sin^3\ttheta}{(1 + \cos^2\ttheta)^2}
 \;,
\label{dY}
\end{equation}
where $\gamma_{\hr}$ and $\gamma_{\ttheta}$ are smooth sphere
function of $(\hr,\ttheta)$ with $\gamma_{\ttheta}(0,\ttheta)
\equiv 1$.

By~\cite[Section 6]{ChCo}, in a sufficiently regular black hole
space-time, in a collar neighbourhood of every component of the
Killing horizon the axis of rotation $\mcA$ has exactly two
connected components, each of which meets a cross section of
the horizon at exactly one point. Now, by Proposition \ref{NHMetric}, in
a neighbourhood of the horizon, $\partial_{\tvarphi}$ vanishes
along $\{\ttheta = 0\}$ and $\{\ttheta = \pi\}$. Evidently
these two sets correspond to different component of $\mcA$.
Denote by $\mcA_+$ and $\mcA_-$ the components of $\mcA$ that
contain $\{\ttheta = 0\}$ and $\{\ttheta = \pi\}$,
respectively.

It is well-known that $Y$ is constant on each component of
$\mcA$. In a neighbourhood of the horizon, this can be seen
readily from the first equation in \eqref{dY}. Away from the
horizon, see e.g.~\cite[Eq.~(2.6)]{CLW} or~\cite{Weinstein1}.
By \eqref{dY}, we have
\[
Y\big|_{\mcA_-} - Y\big|_{\mcA_+}
 = \int_0^\pi Y_{,\ttheta}(0,\ttheta)\,d\ttheta
  = 4\,r_0^2\,\int_0^\pi \frac{\sin^3\ttheta}{(1 + \cos^2\ttheta)^2}\,d\ttheta = 4\,r_0^2
	\;.
\]
Hence, shifting  $Y$ by a constant if necessary, we can assume
that
\begin{equation}
Y\big|_{\mcA_-} = 2r_0^2
	\;, \qquad
Y\big|_{\mcA_+} = -2r_0^2
	\;.
\label{Yaxis}
\end{equation}
Then, by integrating \eqref{dY},
\[
Y = -\frac{4r_0^2\,\cos\ttheta}{1 + \cos^2\ttheta} + \delta Y(\hr,\ttheta)
	\;,
\]
where $\delta Y$ is given by
\begin{align*}
\delta Y(\hr,\ttheta)
	&= 4r_0^2\,\int_{0}^{\ttheta} (\gamma_{\ttheta}(\hr,\tau) - 1)\,\frac{\sin^3 \tau}{(1 + \cos^2\tau)^2}\,d\tau\\
 	&= - 4r_0^2\,\int_{\ttheta}^\pi (\gamma_{\ttheta}(\hr,\tau) - 1)\,\frac{\sin^3 \tau}{(1 + \cos^2\tau)^2}\,d\tau
	\;.
\end{align*}
It thus follows that
\begin{equation}
Y = -\frac{4r_0^2\,\cos\ttheta}{1 + \cos^2\ttheta} + O(\hr\,\sin^4\ttheta)
	\;.
\label{YAsym}
\end{equation}

To proceed, we recall that the distance $d_b$ between two
points $(X_1,Y_1)$ and $(X_2,Y_2)$ in the (real) hyperbolic
plane is implicitly given by the
formula~\cite[Theorem~7.2.1]{Beardon}:
\beaa
    \cosh\,d_b -1 &  = & \frac 12   \bigg(\bigg(\sqrt{ \frac {X_1}{X_2} }-\sqrt{ \frac {X_2}{X_1} } \bigg)^2 +
    \frac{(Y_1-Y_2)^2}{X_1X_2}\bigg)
 \;.
\eeaa
Also, recall that we have shown that the functions $z$ and
$\rho$ defined in Section \ref{SSorbit} provide global
isothermal coordinates on the orbit space. Define $(r,\theta)$
by
\[
(z,\rho) = (r\,\cos\theta, r\,\sin\theta).
\]
Now consider a reference  Ernst potential $\xke + i\,\yke$ as
given in~\cite{Dain:variational}:
\begin{align}
 \label{omval-x}
\xke(r,\theta) &= \Big(\frac{1}{2}(r\sqrt{2} + r_0)^2 + \frac{r_0^2}{2}
+ \frac{r_0^3\,(r\sqrt{2} + r_0)\,\sin^2\theta}{(r\sqrt{2} + r_0)^2 + r_0^2\,\cos^2\theta}\Big)\sin^2\theta
	\;,\\
 \label{omval-y}
\yke(r,\theta) &= r_0^2(\cos^3\theta -3\cos\theta )- \frac{r_0^4\,\cos\theta\,
\sin^4\theta }{(r\sqrt{2} + r_0)^2 + r_0^2\,\cos^2\theta}
	\;.
\end{align}
Here $r$ and $\theta$ are polar coordinates associated to
Kerr's own $(z,\rho)$ coordinates.
It is convenient to rewrite $\xke$ and $\yke$ as
\begin{align}
 \label{omval-x-v}
 \xke(r,\theta) &= \frac{2r_0^2\,\sin^2\theta}{1 + \cos^2\theta} + O(r\,\sin^2\theta)
	\;,\\
\yke(r,\theta) &= -\frac{4r_0^2\,\cos\theta}{1 + \cos^2\theta} + O(r\,\sin^4\theta)
	\;. \label{omval-y-v}
\end{align}
The leading order term near $r=0$ for $\yke$ can also be
rewritten in the following form
\beaa -\frac{4r_0^2\,\cos\theta}{1 + \cos^2\theta} &=& -2r_0^2
+
 \displaystyle
\frac{  r_0^2\, \sin ^4\theta}{2(1+\cos^2
\theta)\cos^4(\theta/2)}
 	\;,
 \displaystyle\label{omval-y-v2}
\eeaa
useful away from $\theta=\pi$, or as
\beaa -\frac{4r_0^2\,\cos\theta}{1 + \cos^2\theta}
 &=&
  2r_0^2 -
 \displaystyle
\frac{  r_0^2\, \sin ^4\theta}{2(1+\cos^2 \theta)\sin^4( \theta
/2)}
 	\;,
 \label{omval-y-v3}
\eeaa
which is useful away from $\theta=0$. This shows that in either
case the deviation from the constant terms $\pm 2r_0^2$ in
$\yke$ factors out through $\sin^4 \theta$.

In the remainder of this section  we  derive a bound for the
hyperbolic distance between $(X,Y)$ and $(\xke,\yke)$, which
are compared after identifying the $(z,\rho)$ coordinates of
the solution under consideration with the $(z,\rho)$
coordinates of the reference Kerr solution. This leads to
relations between $(r,\theta)$ and $(\hr,\ttheta)$, which we
analyze now. By \eq{rho-Asymp} and \eq{z-Asymp} we have
\[
r\,\sin\theta = \rho = \hr\,\tilde\beta_{\rho}\,\sin\ttheta	\;,
	\text{ and } r\,\cos\theta = z = \hr\,\tilde\beta_{z}\,\cos\ttheta	\;.
\]
Thus, for small $\hr$,
\begin{align}
r^2
	&= \rho^2 + z^2 = \hr^2 + O(\hr^3)	\;,
\label{r-Asymp}\\
\sin\theta
	&= (1 + O(\hr))\,\sin\ttheta	\;.
\label{sint-Asymp}\\
\cos\theta
	&= (1 + O(\hr))\,\cos\ttheta	\;.
\label{cost-Asymp}
\end{align}
Substituting \eqref{r-Asymp}-\eqref{cost-Asymp} into \eqref{omval-x-v}-\eqref{omval-y-v} we get
\begin{align}
 \label{xke-Asymp}
\xke(r,\theta) &= \frac{2r_0^2\,\sin^2\ttheta}{1 + \cos^2\ttheta}(1 + O(\hr))
	\;,\\
 \label{yke-Asymp-w}
\yke(r,\theta) &= -\frac{4r_0^2\,\cos\ttheta}{1 + \cos^2\ttheta} + O(\hr \sin^4 \ttheta)
	\;.
\end{align}

From \eqref{X-Ftrztn}, \eqref{YAsym}, \eqref{xke-Asymp} and
\eqref{yke-Asymp-w}  we arrive at
\begin{equation}
d_b((X,Y),(\xke,\yke)) = O(1) \text{ for small } \hr
	\;.
\label{BoundedDistance}
\end{equation}

We have therefore proved (for terminology, see~\cite{ChCo}):

\begin{Theorem}
 \label{Tfd}
Let $(\mcM,g)$ be a vacuum, $I^+$ regular, stationary and
axisymmetric asymptotically flat black hole space-time. Let
$\mcE_0$ be a degenerate component of the event horizon
$I^+(\Mext)\cap \pdoc$ with cross-section  area $A_0$. There
exists a neighbourhood of $\mcE_0$ on which the
hyperbolic-plane distance between the complex Ernst potential
of $(\mcM,g)$ and that of the extreme Kerr space-time with the
same  area of the cross-sections of the horizon is bounded.
\end{Theorem}

\subsection{The Ernst potential of $\partial_{\tvarphi}$ in electrovacuum}
 \label{ssEptev}

We continue with the extension of the analysis in
Section~\ref{ssEptv} to the electrovacuum case. Let $\CF$ be
the electro-magnetic two-form in a stationary axisymmetric
space-time $(\mcM,g)$ satisfying the sourceless
Einstein-Maxwell equations: thus $\CF$ is invariant under both
$\xi$ and $\eta$ and satisfies the Maxwell equations:
\begin{equation}
\CF = d\CA
	\;,\qquad d*\CF = 0
	\;.
\label{maxwell}
\end{equation}

To fix terminology, the \emph{vectorial Ernst potential}
$(U,V,\chie,\chim)$ of the rotational Killing field $\Yfield$,
is defined as follows. First, $U$ is defined as
\[
U = -\frac{1}{2}\log X = -\frac{1}{2}\log g(\Yfield,\Yfield)
	\;.
\]
Next, the electric and magnetic potentials $\chie$ and $\chim$ of $\eta$ are defined by
\[
d\chie
	= i_{\Yfield} *\CF
	\;, \qquad d\chim = i_{\Yfield} \CF
	\;.
\]
To dispel confusion, we emphasize that these are not the same
as the standard electric and magnetic potentials which are
defined using the stationary Killing field. The existence of
$\chie$ and $\chim$ is a consequence of \eqref{maxwell}. Note
that $\Yfield$ vanishes on the axis $\mcA$, which implies that
$\chie$ and $\chim$ are constant on each connected component of
$\mcA$. It further follows from the Einstein-Maxwell equations
that the $1$-form $*(\eta \wedge d\eta) - 2\chim\,d\chie +
2\chie\,d\chim$ is closed (see e.g.~\cite{Weinstein3}), and so
we can define $V$ by
\[
2\,dV \equiv dY := *(\eta \wedge d\eta) - 2\chim\,d\chie + 2\chie\,d\chim
	\;.
\]
Similarly to the vacuum case, $V$ is constant on each connected
component of $\mcA$.

 In the sequel, we analyze the asymptotic behaviour of
$(U,V,\chie,\chim)$ as $\hr\rightarrow0$. It is desired to relate this potential to that of the reference
degenerate Kerr-Newman solution which has the same horizon area
$A_0$ and charge parameters $q_e$ and $q_b$. To this end, we
introduce the variable $(r,\theta)$ as in Section \ref{ssEptv}
by
\[
(z,\rho) = (r\cos\theta, r\sin\theta)
	\;,
\]
where the functions $z$ and $\rho$ are defined in
Section~\ref{SSorbit}.  The Ernst potential of the reference
Kerr-Newman solution then takes the following form in terms of
$r$ and $\theta)$ (see, e.g., \cite{Heusler:book}):
\begin{align*}
\uke
	&= -\frac{1}{2}\log \frac{ \sin^2\theta((r + \mzero)^2 + \azero^2)^2 - r^2\,\azero^2\,\sin^2\theta) }{ (r + \mzero)^2 + a^2\,\cos^2\theta}
	\;,\\
\vke
	&= -\azero\,\mzero\,(3\cos\theta - \cos^3\theta) - \azero\,\mzero\,\cos\theta\,\sin^2\theta\,\frac{\azero^2\,\sin^2\theta - q_0^2\,\mzero^{-1}\,(r + \mzero)}{(r + \mzero)^2 + \azero^2\,\cos^2\theta}
	\;,\\
\chieke
	&= -q_e\,\cos\theta\frac{(r + \mzero)^2 + \azero^2}{(r + \mzero)^2 + \azero^2\,\cos^2\theta} + q_b\,\azero\,\sin^2\theta\,\frac{r + \mzero}{(r + \mzero)^2 + \azero^2\,\cos^2\theta}
	\;,\\
\chimke
	&= -q_b\,\cos\theta\frac{(r + \mzero)^2 + \azero^2}{(r + \mzero)^2 + \azero^2\,\cos^2\theta} - q_e\,\azero\,\sin^2\theta\,\frac{r + \mzero}{(r + \mzero)^2 + \azero^2\,\cos^2\theta}
	\;.
\end{align*}
Here $q_0$ is the total charge,
$$
 q_0 := \sqrt{q_e^2 + q_b^2}
	\;.
$$
Note that in \cite{Heusler:book}, only the case of vanishing magnetic charge is considered. The general magnetically charged solution can be obtained from
this by a duality rotation, $\CF \mapsto \cos \lambda \CF + \sin
\lambda
* \CF$, where $\lambda$ is a real constant. Under this transformation the new potentials $\chieke$ and
$\chimke$ are obtained from the old ones by a constant rotation
in the $(\chieke,\chimke)$ plane, while $\uke$ and $\vke$
remain unchanged, whence the above formulae.

As shown in Section~\ref{ssEptv}, we have the relations
\begin{equation}
\sin\theta = (1 + O(\hr))\sin\ttheta\;, \text{ and } \cos\theta = (1 + O(\hr))\cos\ttheta \text{ for small } \hr
	\;,
\label{tt-ev}
\end{equation}
where the error terms are smooth sphere functions. Thus, by a
simple calculation, in a neighbourhood of the horizon we have
\begin{align}
\uke
	&= - \log \sin\ttheta + O(1)
	\;,\label{uke-a}\\
\vke
	&= 2\azero\,\mzero\,s(\ttheta) + O(\sin^2\ttheta)
	\;,\label{vke-a}\\
\chieke
	&= q_e\,s(\ttheta) + O(\sin^2\ttheta)
	\;,\label{ceke-a}\\
\chimke
	&=  q_b\,s(\ttheta) +O(\sin^2\ttheta)
	\;,\label{cmke-a}
\end{align}
where $s$ is some smooth function of $\ttheta$ such that
$s(\ttheta)\equiv-1$ in $[0,\pi/6]$ and $s(\ttheta)\equiv1$ in
$[5\pi/6,\pi]$.

We are ready for the analysis of $(U,V,\chie,\chim)$ near the
horizon. From the flux formulae for the total electric and
magnetic charges of the horizon we have
\begin{align}
 q_b
	&= \frac{1}{2}\int_0^\pi \chim_{,\ttheta}\,d\ttheta
	\;,\label{MagCh}\\
q_e
	&= \frac{1}{2}\int_0^\pi \chie_{,\ttheta}\,d\ttheta
	\;.
\label{ElCh}
\end{align}
In view of the above two identities,
we can assume without loss of generality that
\begin{equation}
\chim\big|_{\mcA_\pm} = \mp q_b
	\;,\qquad	\chie\big|_{\mcA_\pm} = \mp q_e
	\;,
\label{chi-endpt}
\end{equation}
where $\mcA_\pm$ are the connected components of the axis
$\mcA$ as defined in Section~\ref{ssEptv}.

By Proposition~\ref{NHMetric} we have%
\footnote{We use the convention $\CF=\CF_{\mu\nu} dx^\mu \wedge
dx^\nu$ for the coefficients $\CF_{\mu\nu}$ of a two-form.}
\begin{equation}
\chim_{,\ttheta}(\hr,\ttheta) = 2\CF_{\varphi\ttheta}(\hr,\ttheta) = O(\sin\ttheta) \text{ for small } \hr
	\;,
\label{chimt-Asymp}
\end{equation}
and so, by \eqref{chi-endpt}, again for small  $\hr$,
\begin{equation}
\chim(\hr,\ttheta) = -q_b + \int_0^{\ttheta} \chim_{,\ttheta}(\hr,\ttheta)d\tau
= q_b -\int_{\ttheta}^\pi \chim_{,\ttheta}(\hr,\ttheta) = q_b\,s(\ttheta) + O(\sin^2\ttheta)
	\;.
\label{chim-Asymp}
\end{equation}
Similarly, we have
\begin{equation}
\chie_{,\ttheta}(\hr,\ttheta) = 2*\CF_{\varphi\ttheta}(\hr,\ttheta) = O(\sin\ttheta) \text{ for small } \hr
	\;,
\label{chiet-Asymp}
\end{equation}
leading to
\begin{equation}
\chie(\hr,\ttheta) = q_e\,s(\ttheta) + O(\sin^2\ttheta) \text{ for small } \hr
	\;.
\label{chie-Asymp}
\end{equation}

Next, applying the result of~\cite{LP1} to $(\mcM,g,\CF)$ and
to $(\mcM,g,*\CF)$ we find
\[
\CF_{\varphi\ttheta}(0,\ttheta) = \CF_{\mbox{\scriptsize KN}\varphi\ttheta}(0,\ttheta)
	\;,\text{ and } *\CF_{\varphi\ttheta}(0,\ttheta)
    = *\CF_{\mbox{\scriptsize KN}\varphi\ttheta}(0,\ttheta)
	\;,
\]
where $\CF_{\mbox{\scriptsize KN}}$ is the electro-magnetic
two-form associated to the reference Kerr-Newman solution. Note
that here we have used $\theta\equiv\ttheta$ on the horizon. It
thus follows from the definition of $\chie$ and $\chim$ that
\[
\chie(0,\ttheta) \equiv \chieke(0,\ttheta)
	\;, \text{ and } \chim(0,\ttheta) \equiv \chimke(0,\ttheta)
	\;.
\]
By a direct computation, we then get
\begin{equation}
 \int_0^\pi \big[-\chim\,\chie_{,\ttheta} + \chie\,\chim_{,\ttheta}\big]_{\hr = 0}\,d\ttheta
 = \frac{q_0^2\,r_0^4}{\azero^2\,\mzero^2}\,\arctan\frac{\azero}{\mzero} - \frac{q_0^4}{\azero\,\mzero}
	\;.
\label{Vgap1}
\end{equation}
To continue, we recall \eqref{twist} which gives
\[
*(\Yfield \wedge d\Yfield)_{\ttheta} = h_{\tvarphi\tvarphi}\,\psi^{-1}\,\sqrt{\det h}\,[\hr\,\alpha]_{,\hr}
	\;.
\]
Hence, by \eqref{r=0-psi-ev}, \eqref{r=0-alpha-ev} and \eqref{KeyFtrztn-ev},
\begin{equation}
*(\Yfield \wedge d\Yfield)_{\ttheta} = \frac{2\azero\,\mzero\,r_0^4\,\sin^3\ttheta}{(\mzero^2 + \azero^2\,\cos^2\ttheta)^2}\,\gamma_{\ttheta}
	\;,
\label{twistt-Asymp}
\end{equation}
where $\gamma_{\ttheta}$ is a smooth sphere function satisfying
$\gamma_{\ttheta}(0,\ttheta) \equiv 1$. It follows that
\begin{equation}
\int_0^\pi *(\Yfield \wedge d\Yfield)_{\ttheta}\big|_{\hr = 0}\,d\ttheta
= -\frac{2\,q_0^2\,r_0^4}{\azero^2\,\mzero^2}\,\arctan\frac{\azero}{\mzero} + 2\frac{r_0^4}{\azero\,\mzero}
	\;.
\label{Vgap2}
\end{equation}
Using \eqref{Vgap1}, \eqref{Vgap2} and recalling the definition
of $V$ we get
\begin{equation}
V\big|_{\mcA_-} - V\big|_{\mcA_+} = \int_0^\pi V_{,\ttheta}(0,\ttheta)\,d\ttheta = \frac{r_0^4}{\azero\,\mzero} - \frac{q_0^4}{\azero\,\mzero} = 4\,\azero\,\mzero
	\;.
\label{Vgap}
\end{equation}
Thus, we can assume without loss of generality that
\[
V\big|_{\mcA_\pm} = \mp 2\,\azero\,\mzero
	\;.
\]

Now, taking \eqref{chimt-Asymp}, \eqref{chim-Asymp},
\eqref{chiet-Asymp}, \eqref{chie-Asymp} and
\eqref{twistt-Asymp} into account we get
\[
V_{,\ttheta}(\hr,\ttheta) = O(\sin\ttheta) \text{ for small } \hr
	\;.
\]
Thus
\begin{equation}
V(\hr,\ttheta) = 2\azero\,\mzero\,s(\ttheta) + O(\sin^2\ttheta) \text{ for small } \hr
	\;.
\label{V-Asymp}
\end{equation}
Finally, by \eqref{KeyFtrztn-ev},
\begin{equation}
U(\hr,\ttheta) = - \log \sin\ttheta + O(1) \text{ for small } \hr
	\;.
\label{U-Asymp}
\end{equation}

Recall that the distance $d_b$ between two points
$p_1=(U_1,V_1,{\chie}_{1}, \chim_1)$ and
$p_2=(U_2,V_2,{\chie}_{2}, \chim_2)$ in the complex hyperbolic
plane is given by (see, e.g., \cite[Equation~(55),
p.~26]{WeinsteinDuke})
\begin{multline*}
\cosh^2 d_b(p_1, p_2)
	= \Big\{ \cosh(U_1 - U_2) + e^{U_1 + U_2}\big[(\chie_1 - \chie_2)^2 + (\chim_1 - \chim_2)^2\Big\}^2 \\
		+ e^{2(U_1 + U_2)}\Big\{V_1 - V_2 - \chim_2(\chie_1 - \chie_2) + \chie_2(\chim_1 - \chim_2)\Big\}^2
	\;.
\end{multline*}
We conclude from \eqref{U-Asymp}, \eqref{V-Asymp}, \eqref{chie-Asymp}, \eqref{chim-Asymp} and \eqref{uke-a}-\eqref{cmke-a} that
\begin{equation}
d_b((U,V,\chie,\chim),(\uke,\vke,\chieke,\chimke)) = O(1) \text{ for small } \hr
	\;.
\label{BdedDist-ev}
\end{equation}

We have therefore proved (for terminology, see~\cite{ChCo}):

\begin{Theorem}
 \label{Tfd-ev}
Let $(\mcM,g, \CF)$ be an electrovacuum, $I^+$ regular,
stationary and axisymmetric asymptotically flat black hole
space-time. Let $\mcE_0$ be a degenerate component of the event
horizon $I^+(\Mext)\cap \pdoc$ with cross-section area $A_0$,
electric charge $q_e$ and magnetic charge $q_b$. There exists a
neighbourhood of $\mcE_0$ on which the complex-hyperbolic-plane
distance between the vectorial Ernst potential of the
rotational Killing vector field of $(\mcM,g)$ and that of the
extreme Kerr-Newman space-time with the same horizon
cross-section area, electric charge and magnetic charge is
bounded.
\end{Theorem}

\begin{Remark}
A more careful analysis as in Section~\ref{ssEptv} shows the
following asymptotic behaviour for small $\hr$:
\begin{align*}
|\chie - \chieke| + |\chim - \chimke| + |V - \vke|
	&\le  C \hr\,\sin^2\ttheta
	\;,\\
|V - \vke - \chimke(\chie - \chieke) + \chieke(\chim - \chimke)|
	&\le  C \hr\,\sin^4\ttheta
	\;.
\end{align*}
\end{Remark}

\section{Proof of Theorem \ref{Tmain30I.1}}
 \label{Sproof}

Let $(\mcM,g)$ be a stationary, $I^+$-regular, analytic
electrovacuum space-time with connected, non-empty, rotating
future event horizon $\mcE_0$. As justified in detail
in~\cite{ChCo}, we only need to consider the case where the
metric is axisymmetric. By Theorem \ref{T30I.1}, the area
function $\rho$ and its harmonic conjugate $-z$ form a global
manifestly asymptotically flat coordinate system on
$\doc/(\R\times\Uone)$, where $\mcE_0$ corresponds to the point
$\rho = z = 0$. It is well known that the vectorial Ernst
potential $(U,V,\chie,\chim)$ is a harmonic map from
$\RR^3\setminus \{\rho = 0\} = \{(\rho,z,\tvarphi: \rho > 0, z
\in \RR, \tvarphi \in [0,2\pi]\}$ into the complex hyperbolic
plane. Define a reference vectorial Ernst potential
$(\uke,\vke,\chieke,\chimke)$ as in Section \ref{ssEptev}. By
the asymptotic analysis of~\cite{SimonBeig} (compare
\cite{Costaelvac}) and Theorem \ref{Tfd-ev}, the hyperbolic
distance $d_b$ between the two Ernst potentials is finite and
goes to zero as one recedes to infinity. Using the
subharmonicity of $d_b$ and~\cite[Proposition C.4]{CLW}, we
conclude that $d_b \equiv 0$ and so $(U,V,\chie,\chim) \equiv
(\uke,\vke,\chieke,\chimke)$. It is then customary to show that
$(\Mext,g)$ is diffeomorphic to the corresponding domain of
outer communications in that Kerr-Newman space-time to which
the reference Ernst potential is associated.\eproof


\bigskip
\noindent{\textsc{Acknowledgements:} Both authors were
supported in part by the EPSRC Science and Innovation award to
the Oxford Centre for Nonlinear PDE (EP/E035027/1).  PTC was
further supported in part by the Polish Ministry of Science and
Higher Education grant Nr N N201 372736. LN would like to thank Dr. Willie W.Y. Wong for drawing his attention to the problem studied in the present paper.}

\bibliographystyle{amsplain}
\bibliography{../references/hip_bib,%
../references/reffile,%
../references/newbiblio,%
../references/newbiblio2,%
../references/bibl,%
../references/howard,%
../references/bartnik,%
../references/myGR,%
../references/newbib,%
../references/Energy,%
../references/netbiblio}

\end{document}